\documentclass[aps, reprint, floatfix,
superscriptaddress, pra, longbibliography]{revtex4-1}
\usepackage{blindtext}
\usepackage{braket}
\usepackage{graphicx} 
\usepackage[T1]{fontenc}
\usepackage{amsmath,amsthm,amsfonts}
\usepackage{bm}
\usepackage{varioref}
\usepackage{dsfont}
\usepackage[colorlinks=true,linkcolor=blue,citecolor=blue,filecolor=black, urlcolor=blue]{hyperref}
\usepackage{xr-hyper}
\usepackage{cleveref}
\usepackage{tikz}
\newcommand{\mycomment}[1]{}
\usetikzlibrary{arrows.meta, positioning, calc, fit}
\newcommand{\e}{\mathrm{e}}

\renewcommand{\Re}{\mathrm{Re}}
\renewcommand{\Im}{\mathrm{Im}}

\begin{document}
\title{Heisenberg-scaling characterization of a two-channel optical network via two-port homodyne detection}
\author{Atmadev Rai}
\affiliation{School of Mathematics and Physics, University of Portsmouth, Portsmouth PO1 3QL, United Kingdom}
\affiliation{Quantum Science and Technology Hub, University of Portsmouth, Portsmouth PO1 3QL, United Kingdom}
\author{Paolo Facchi}
\affiliation{Dipartimento di Fisica, Universit\`{a} di Bari \textup{\&} Politecnico di Bari, I-70126 Bari, Italy}
\affiliation{INFN, Sezione di Bari, I-70126 Bari, Italy}
\author{Vincenzo Tamma}
\email[]{vincenzo.tamma@port.ac.uk}
\affiliation{School of Mathematics and Physics, University of Portsmouth, Portsmouth PO1 3QL, United Kingdom}
\affiliation{Quantum Science and Technology Hub, University of Portsmouth, Portsmouth PO1 3QL, United Kingdom}
\affiliation{Institute of Cosmology and Gravitation, University of Portsmouth, Portsmouth PO1 3FX, United Kingdom}
\begin{abstract}
We present a fully Gaussian and experimentally feasible scheme for the simultaneous estimation of the four real parameters that characterize a two-channel optical network. The scheme utilizes a two-mode squeezed probe and balanced homodyne detection at both output ports, for which we derive the complete classical Fisher information matrix analytically. Our scheme achieves the Heisenberg-scaling sensitivity for all four parameters simultaneously, enabling full multiparameter characterization of the two-channel interferometric network. We further show, by maximum-likelihood estimation, that the corresponding multiparameter Cram\'er-Rao bounds are saturated with a modest number of experimental repetitions and for low photon number. The scheme establishes a practical route to Heisenberg-scaling multiparameter Gaussian metrology for a two-channel network, with direct relevance to calibration and sensing in integrated photonics and distributed quantum-enhanced measurement architectures.
\end{abstract}

\maketitle
Extracting multiple unknown physical quantities from the same experiment is rapidly becoming a cornerstone requirement for quantum sensing. Many realistic sensing tasks are inherently multiparametric, including quantum imaging and phase-based schemes, which require the simultaneous estimation of multiple phases~\cite{PhysRevLett.111.070403,PhysRevA.94.042342,albarelli2020perspective}, distributed sensing protocols that infer multiple phases or field components across spatially separated sensor networks~\cite{PhysRevA.97.032329,PhysRevLett.120.080501,guo2020distributed}, and quantum process tomography used for device characterization~\cite{zhou2015quantum,PhysRevA.100.012350}. These examples highlight the need for a multiparameter framework for quantum  metrology~\cite{szczykulska2016multi,ALBARELLI2020126311}. In the single-parameter setting, quantum-enhanced metrology often surpasses the classical shot-noise limit by exploiting nonclassical resources such as squeezing and entanglement, and can in principle achieve the so-called Heisenberg-limit sensitivity~\cite{PhysRevD.23.1693,PhysRevLett.71.1355,PhysRevD.30.2548,PhysRevLett.96.010401,dowling2015quantum,zhou2018achieving,qian2019heisenberg,doi:10.1126/science.1104149,PhysRevLett.102.100401,PhysRevA.92.022106}.

Extending single-parameter protocols to the multiparameter setting is, however, not straightforward. In general, multiparameter estimation is constrained by trade-offs arising from probe incompatibility, where a probe optimal for one parameter need not be optimal for all parameters, and the measurement incompatibility, where the measurements individually optimal for different parameters may not be jointly implementable~\cite{ALBARELLI2020126311,PhysRevX.12.011039,candeloro2021quantum, PhysRevA.94.052108}. Achieving Heisenberg-limited precision and saturating the quantum Cram\'er-Rao bounds (CRBs) simultaneously for multiple parameters may therefore not be possible in general due to these trade-offs. However, a more practical objective from an experimental point of view is to identify scalable probe states and feasible measurements that nevertheless achieve Heisenberg precision scaling as $O(1/N)$ (with $N$ denoting the mean photon number) simultaneously across the entire parameter space of a multiparameter estimation problem~\cite{PhysRevResearch.3.013152, PhysRevA.111.062408}. 

Continuous-variable quantum optics provides a natural framework for this task, and Gaussian states are especially well suited because they provide both experimental robustness and a compact theoretical description~\cite{ferraro2005gaussian, RevModPhys.77.513,Adesso_2007}. Their statistics are fully characterized by first and second moments of the field quadratures, covering coherent and thermal states as well as squeezed and entangled resources. The Gaussian framework also supports robust estimation methods that perform well under finite sampling and experimental noise, which is essential when the aim is not only asymptotic scaling but also experimental feasibility.
\begin{figure}
    \centering
    \includegraphics[width=\linewidth]{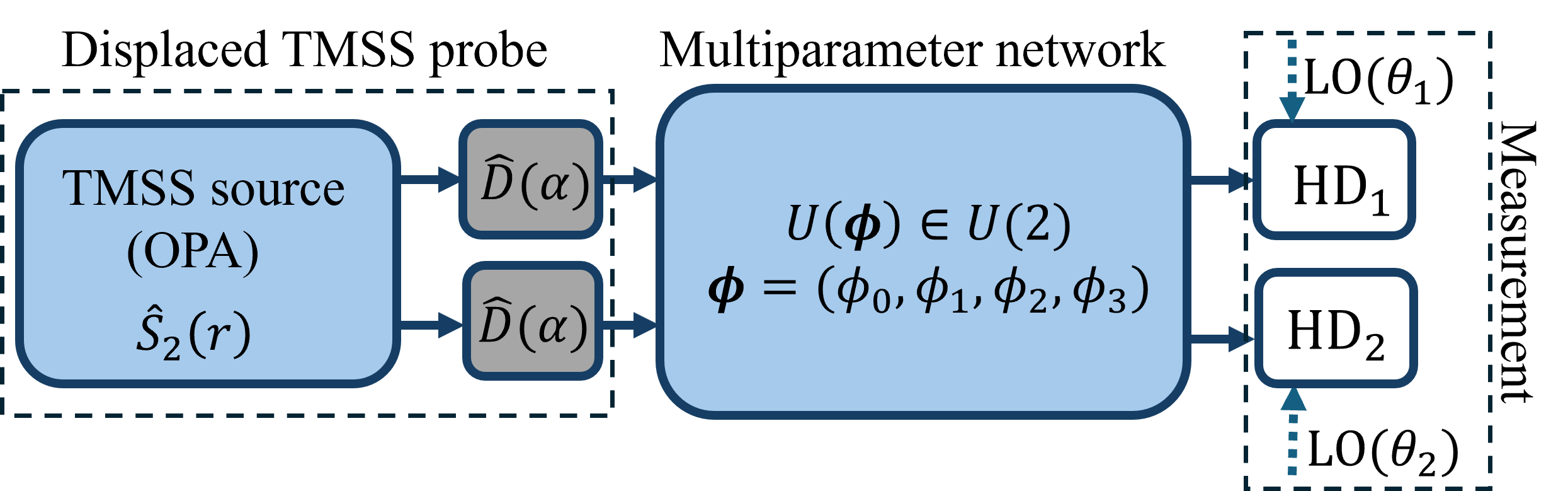}
   \caption{A two-mode squeezed state (TMSS) (generated, for example, by an optical parametric amplifier (OPA)) probes an unknown two-channel linear-optical network described by an arbitrary unitary \(U(\bm{\phi})\) in Eq.~\eqref{Eq:Unitary}, with parameter vector $\bm{\phi}=(\phi_0,\phi_1,\phi_2,\phi_3)$, where \(\phi_3\) sets the mode mixing while \(\phi_1,\phi_2\) and \(\phi_0\) represent internal and overall phase parameters (the latter becomes operationally identifiable in the presence of a phase reference such as a local oscillator (LO)). The two output modes are measured via balanced homodyne detection (HD) at both ports, using tunable LO phases \(\theta_1\) and \(\theta_2\).}
\label{fig:scheme_U2_homodyne}
\end{figure}

In a two-channel setting, the most general passive linear-optical device is described by a $\mathrm{U}(2)$ transformation, which is fully specified by four real parameters and spans the complete family of two-mode operations, including phase shifters and mode-mixing angles~\cite{PhysRevA.40.1371,macho2021optical}. Establishing quantum-enhanced, multiparameter estimation strategies in this setting is therefore directly relevant to precision sensing and to full characterization of photonic devices.

In this Letter, we present an experimentally feasible scheme for simultaneously estimating the four real parameters of a two-channel linear-optical network. Within the Gaussian framework, we consider an experimentally accessible two-mode squeezed state (TMSS) with equal displacement amplitudes injected into each input channel and perform a two-port balanced homodyne detection at the output channels of the linear network. In the scheme, we experimentally tune the local oscillator phases of the homodyne detection near the quadratures with the minimum variance using the coarse measurement outcomes. The scheme achieves Heisenberg scaling for all three phase parameters across their full domain, while the mode-mixing angle attains the Heisenberg scaling only near the balanced working point $\pi/4$. We derive the full Fisher-information matrix (FIM) and evaluate the analytical forms of the corresponding CRBs for the four-parameter estimation problem simultaneously. Furthermore, we identify how the allocation of resources between displacement and squeezing shapes the attainable sensitivities. The present work applies to the estimation of a full set of parameters in a two-channel optical network beyond specific interferometric architectures such as Mach-Zehnder-type networks~\cite{PhysRevA.111.062408}. Furthermore, given that the scaling in the number of resources in the quantum CRBs is not guaranteed, particularly in multiparameter metrology~\cite{537s-r3t8}. We demonstrate the practical attainability of the multiparameter Heisenberg scaling by exploiting the experimentally tunable homodyne measurements. In particular, we showed such scaling within associated classical CRBs and their saturation through a maximum-likelihood analysis. Within the operating regime, our scheme suffices to achieve the Heisenberg scaling simultaneously for all four parameters in a two-channel optical network as described in Fig.~\ref{fig:scheme_U2_homodyne}.

\paragraph*{Quantum sensing scheme for an arbitrary unknown linear network---}
\label{sec:model}
We describe here the quantum sensing scheme in Fig.~\ref{fig:scheme_U2_homodyne} for an arbitrary, lossless two-channel linear-optical network acting on two spatial modes with annihilation operators $\hat a_1$ and $\hat a_2$. These obey the canonical bosonic commutation relations \([\hat a_j,\hat a_k^\dagger]=\delta_{jk}\). In the absence of loss and gain, the evolution is passive and preserves the total photon number, and is therefore fully described by a unitary matrix \(U\in \mathrm{U}(2)\) that acts on the mode operators as
\begin{equation}
\begin{pmatrix}
\hat a_{1,\mathrm{out}}\\[2pt]
\hat a_{2,\mathrm{out}}
\end{pmatrix}
=
U
\begin{pmatrix}
\hat a_{1,\mathrm{in}}\\[2pt]
\hat a_{2,\mathrm{in}}
\end{pmatrix}.
\end{equation}
We employ the convenient parameterization
\begin{equation}
U
=
e^{-i\phi_0}
\begin{pmatrix}
e^{i\frac{\phi_{\tau}}{2}}\cos(\phi_3)
&
e^{i\frac{\phi_{\rho}}{2}}\sin(\phi_3)
\\
-\,e^{-i\frac{\phi_{\rho}}{2}}\sin(\phi_3)
&
e^{-i\frac{\phi_{\tau}}{2}}\cos(\phi_3)
\end{pmatrix},
\label{Eq:Unitary}
\end{equation}
where $\phi_\tau=(\phi_1+\phi_2)/2$ and $\phi_\rho=(\phi_1-\phi_2)/2$ denote the phase associated with transmitted and reflected amplitudes, respectively. The mode-mixing angle \(\phi_3\in[0,\pi/2]\) fixes the transmittance and reflectance,
$\tau=\cos^{2}(\phi_3)$ and $\rho=\sin^{2}(\phi_3)$, respectively, satisfying $\tau+\rho=1$ for a lossless device. The prefactor $e^{-i\phi_0}$ fixes the global phase (and thus $\mathrm{\det}(U)=\e^{-2i\phi_0}$), while $(\phi_1,\phi_2,\phi_3)$ span an arbitrary $\mathrm{SU}(2)$ transformation, corresponding to internal phase shifts $\phi_1,\phi_2\in[0,2\pi]$. 

To probe the arbitrary two-channel unitary, we employ a displaced two-mode squeezed Gaussian probe,
\begin{equation}
\ket{\psi_{\mathrm{in}}}
=\hat D_1(\alpha_1)\,\hat D_2(\alpha_2)\,\hat S_{12}(\xi)\,\ket{0,0},
\label{eq:INPUTstate}
\end{equation}
where \(\hat D_j(\alpha_j)=\exp(\alpha_j\hat a_j^\dagger-\alpha_j^*\hat a_j)\) and
\(\hat S_{12}(\xi)=\exp\!\big(\xi\,\hat a_1\hat a_2-\xi^*\,\hat a_1^\dagger\hat a_2^\dagger\big)\).  For simplicity, and because it optimises the quantum Fisher information, we choose equal real displacements, \(\alpha_1=\alpha_2=\alpha \in \mathbb{R}\), and real squeezing,
\(\xi=r>0\)~\cite{537s-r3t8}. We quantify the resources in terms of the mean photon numbers carried by squeezing and displacement. The average number of photons in the squeezing is $N_s=2\sinh^2(r)$, while the total number of displacement photons at the input is $N_c=2 \alpha^2$. The total mean photon number injected into the network is then $N=N_s+N_c$.

The output modes are then measured by balanced homodyne detection at both ports, with tunable LO phases \(\theta_1\) and \(\theta_2\), producing joint Gaussian statistics from which all four parameters are inferred. Note that \(\phi_0\) multiplies the transformation as a global factor \(e^{i\phi_0(\hat{a}_1^\dagger\hat{a}_1+\hat{a}_2^\dagger\hat{a}_2)}\) and is therefore unobservable for number-diagonal probes and POVMs. It becomes operationally identifiable once a phase reference is present, and should therefore be understood as an overall phase defined relative to the reference field. In our scheme, the homodyne detection introduces number coherence relative to the LO, so \(\phi_0\) rotates the output quadratures with respect to the LO and can be estimated~\cite{PhysRevA.91.032103}. In this sense, the four-parameter characterization discussed here refers to a phase-referenced $\rm{U}(2)$ transformation.

\paragraph*{Two-port homodyne detection and Fisher information matrix---}
\label{sec:FIM_main}
Because both the probe state and the measurement are Gaussian, each experimental run yields a pair of real homodyne outcomes \(\bm{x} = (x_1,x_2)^{\mathsf T}\) distributed according to a bivariate Gaussian
distribution with mean vector \(\bm{\mu}_{\bm{\phi}}\) and covariance
matrix \(\Sigma_{\bm{\phi}}\),
\begin{equation}
  p(\bm{x}\,|\,\bm{\phi})
  =
  \frac{1}{2\pi\sqrt{\det\Sigma_{\bm{\phi}}}}\,
  \exp\!\left[
    -\frac{1}{2}
    \big(\bm{x}-\bm{\mu}_{\bm{\phi}}\big)^{\mathsf T}
    \Sigma_{\bm{\phi}}^{-1}
    \big(\bm{x}-\bm{\mu}_{\bm{\phi}}\big)
  \right],
  \label{eq:Gaussian_pdf_main}
\end{equation}
where \(\bm{\phi} = (\phi_0,\phi_1,\phi_2,\phi_3)\) denotes the set of
unknown parameters defining the two–channel network. The explicit expressions for \(\bm{\mu}_{\bm{\phi}}\) and \(\Sigma_{\bm{\phi}}\) are given in  the Supplemental Material (SM)~\cite{Supp}.
\nocite{PhysRevA.96.062304, PRXQuantum.3.010202}

Since all four unknown parameters are encoded simultaneously in the output statistics, the analysis of the precision for the scheme requires a multiparameter estimation framework. In this setting, the attainable precision is governed by the FIM, which quantifies the local sensitivity of the likelihood $p(\bm{x}|\bm{\phi})$ to change in each component of $\bm{\phi}$ and, through the multiparameter CRB, sets the best achievable covariance of any unbiased joint estimator. 

For the Gaussian distributions of the form in Eq.~\eqref{eq:Gaussian_pdf_main}, the classical FIM takes the standard form~\cite{cramer1999mathematical}
\begin{equation}
  F_{ij}
  =
  \frac{1}{2}\,
  \mathrm{Tr}\!\left[
    \Sigma_{\bm{\phi}}^{-1}
    \big(\partial_i \Sigma_{\bm{\phi}}\big)
    \Sigma_{\bm{\phi}}^{-1}
    \big(\partial_j \Sigma_{\bm{\phi}}\big)
  \right]
  +
  \big(\partial_i \bm{\mu}_{\bm{\phi}}\big)^{\mathsf T}
  \Sigma_{\bm{\phi}}^{-1}
  \big(\partial_j \bm{\mu}_{\bm{\phi}}\big),
  \label{eq:FIMGaussian_main}
\end{equation}
where \(\partial_i\equiv\partial/\partial\phi_i\). It is convenient to write the two terms in the FIM as $ F = F^{\Sigma} + F^{\bm{\mu}}$. The term \(F^{\Sigma}\) quantifies the information carried by parameter-dependent fluctuations and thus by squeezing-induced noise, whereas the term \(F^{\bm{\mu}}\) arises from the variation of parameter dependence of the mean vector \(\bm{\mu}_{\bm{\phi}}\) and is therefore associated with the homodyne signal.

\paragraph*{Multiparameter Heisenberg scaling---}
\begin{figure}
    \centering
    \includegraphics[width=\linewidth]{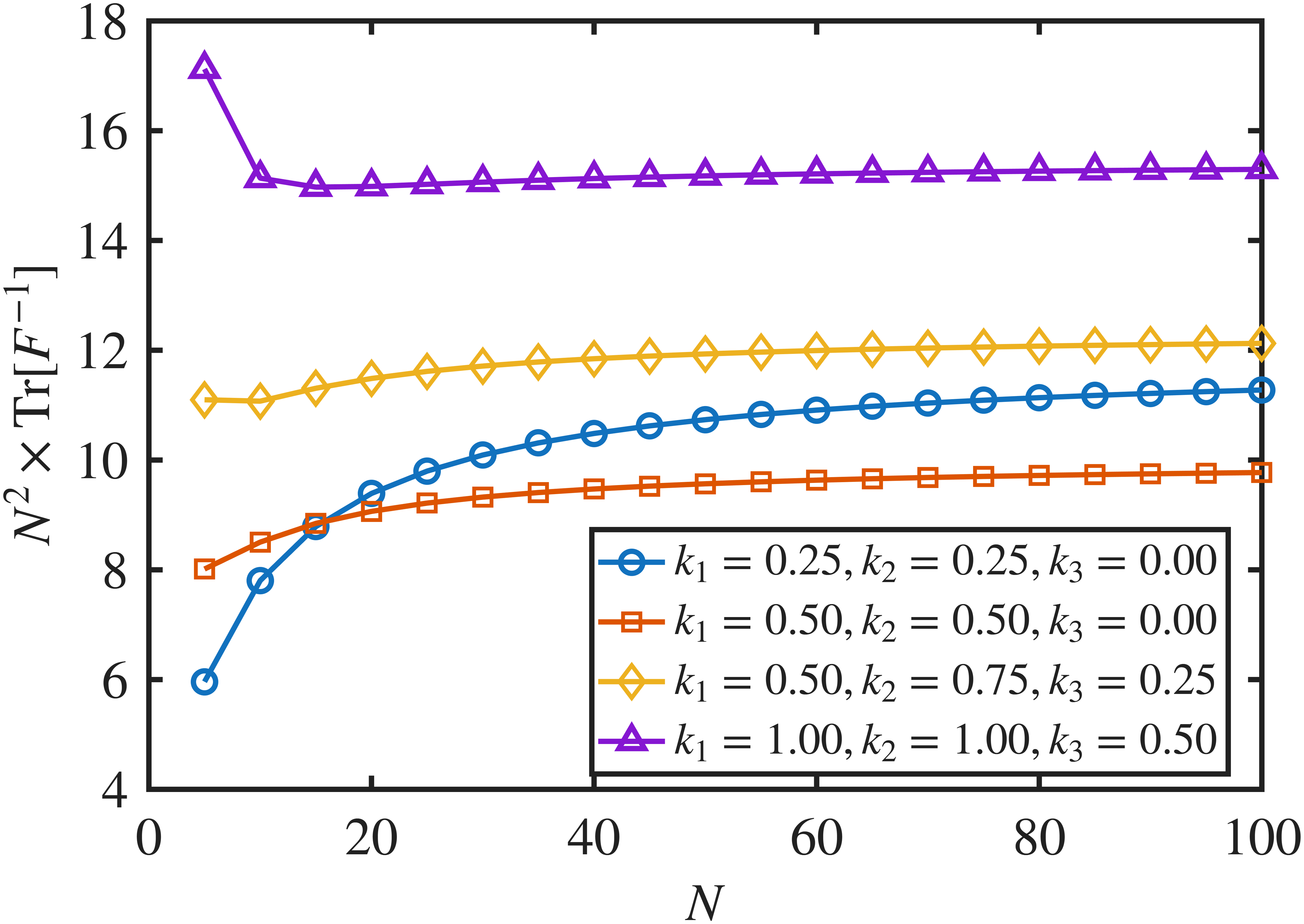}
    \caption{The plot shows the Heisenberg-normalized scalar bound $N^{2}\,\mathrm{Tr}[F^{-1}]$ as a function of the total mean photon number $N$ for several choices of $k_1,k_2,k_3$. Since asymptotically $\mathrm{Tr}[F^{-1}] \propto 1/N^{2}$, the curves approach an $N$-independent plateau, which gives the leading prefactor set by the dominant $N^2$ contribution $\mathcal{F}$ to the Fisher information matrix in Eq.~\eqref{eq:Ftot_asym_main}. Varying $k_1,k_2,k_3$ therefore changes the prefactor while preserving the $1/N^{2}$ scaling. Here $\beta=1/2$.}
    \label{fig:effec_FIM}
\end{figure}
 
We now identify the operating conditions under which the simultaneous estimation of \(\bm{\phi}=(\phi_0,\phi_1,\phi_2,\phi_3)\) exhibits Heisenberg scaling. To this end, we analyze the two terms of the FIM in Eq.~\eqref{eq:FIMGaussian_main}. The explicit forms of $\Sigma_{\bm{\phi}}$ and $\bm{\mu}_{\bm{\phi}}$ reveal an important structural feature of the estimation problem: the covariance matrix \(\Sigma_{\bm{\phi}}\) is independent of the phase \(\phi_1\), whereas the mean vector \(\bm{\mu}_{\bm{\phi}}\) depends on all four parameters. The independence of $\Sigma_{\bm{\phi}}$ from $\phi_1$ follows from the structure of the TMSS probe, as the relative-number fluctuation associated with $\phi_1$ vanishes, i.e.\ $\mathrm{Var}(\hat{N}_1-\hat{N}_2)=0$, where $\hat{N}_1-\hat{N}_2$ denotes the photon-number difference operator between the channels $1$ and $2$, associated with the phase difference $\phi_1$. Consequently, the parameter \(\phi_1\) cannot be inferred from the noise term \(F^{\Sigma}\), its Fisher information enters only through \(F^{\bm{\mu}}\), via the displacement-induced response of the homodyne signal. By contrast, $\phi_0$, $\phi_2$, and $\phi_3$ are encoded in both the mean vector and the covariance matrix~\cite{Supp}.

To make the Heisenberg scaling explicit in an experimentally controlled way, the LO phases can be tuned so that each detector measures close to the minimum-variance quadratures of the corresponding output modes. Specifically, the LO phases follow the asymptotic form
\begin{equation}
\theta_i=f_i+\frac{k_i}{N_s},\qquad i=1,2,
\label{Eq:LOtuning}
\end{equation}
where \(f_i\) denotes the phase of the quadrature fields $\hat{x}_{i,f_i}$ at which the variance of output mode $i$ is minimized. In other words, the LO phases are detuned from the minimum $f_i$ by an additional term $k_i/N_s$ where $k_i$ is an arbitrary constant independent of $N$. For our parameterization, \(f_1=\phi_0+\phi_2/2\) and \(f_2=\pi/2+\phi_0-\phi_2/2\), which can be obtained experimentally by minimizing $\Sigma_{\bm{\phi}}$ from a coarse estimation of the output second moments. We furthermore operate the mode mixing parameter close to the balanced working point,
\(\phi_3=\pi/4+\delta\phi_3\), this is an experimentally relevant situation in which the beam splitter is tuned near a balanced mixing angle, while allowing small deviations. In the asymptotic analysis we take \(\delta\phi_3=k_3/N_s\) with \(k_3\) independent of \(N\).

A detailed derivation of the FIM is given in Sec.~\ref{Appendix B} of the SM~\cite{Supp}. Under the LO-tuning condition~\eqref{Eq:LOtuning} and operating the beam splitter near the balanced point in $\phi_3$, we substitute the derivatives of \(\Sigma_{\bm{\phi}}\) and \(\bm{\mu}_{\bm{\phi}}\) into Eq.~\eqref{eq:FIMGaussian_main} and expand for large photon number $N$ with $N_{s,c}={O}(N)$. The two contributions to the FIM then take the asymptotic forms 
\begin{equation}
F^{\Sigma} = N_{s}^2\,\mathcal{F}^{\Sigma}(k_1,k_2,k_3)
+{O}(N),
\label{eq:AsymFSigma}
\end{equation}
\begin{equation}
F^{\bm{\mu}} = N_{s}N_{c}\,\mathcal{F}^{\bm{\mu}}(k_1,k_2,k_3)
+{O}(N).
\label{eq:AsymFmu}
\end{equation}
Here $\mathcal{F}^{\Sigma}$ and $\mathcal{F}^{\bm{\mu}}$ are $4\times4$ coefficient matrices independent of $N$ and are given in~\cite{Supp}. The term \(F^{\Sigma}\), which arises from the fluctuation of homodyne outcomes, depends solely on squeezing in the probe and scales as \(N_{\mathrm{s}}^{2}\). It therefore gives the leading \(N^{2}\) information for the parameters encoded in the covariance, namely \(\phi_{0}\), \(\phi_2\), and \(\phi_3\). Whereas, $\phi_1$ rotates only the measured quadrature signal in phase space and thus enters only through the homodyne mean $\bm{\mu}_{\bm{\phi}}$. Since $\Sigma_{\bm{\phi}}$ is independent of $\phi_1$, its Fisher information contribution is carried only by \(F^{\bm{\mu}}\), with leading scaling \(N_sN_c\). The coherent displacement is therefore essential for rendering the full four-parameter problem identifiable at Heisenberg scaling.

The physical mechanism behind Eqs.~\eqref{eq:AsymFSigma} and~\eqref{eq:AsymFmu} is the strong anisotropy of the two-mode squeezed noise. 
The $\mathrm{U}(2)$ transformation rotates and mixes the squeezed and anti-squeezed quadrature directions, and homodyne 
detection close to the minimum-variance quadratures converts these small rotations into large relative changes of the 
measured covariance, yielding $F^\Sigma=O(N_s^2)$ for $\phi_0,\phi_2$ and $\phi_3$. By contrast, $\phi_1$ is a blind 
direction of the TMSS covariance, because the photon-number-difference fluctuations vanish. The coherent displacement 
provides the missing phase-sensitive mean field, giving $F^\mu_{\phi_1\phi_1}=O( N_sN_c$).
\begin{figure}
    \centering
    \includegraphics[width=\linewidth]{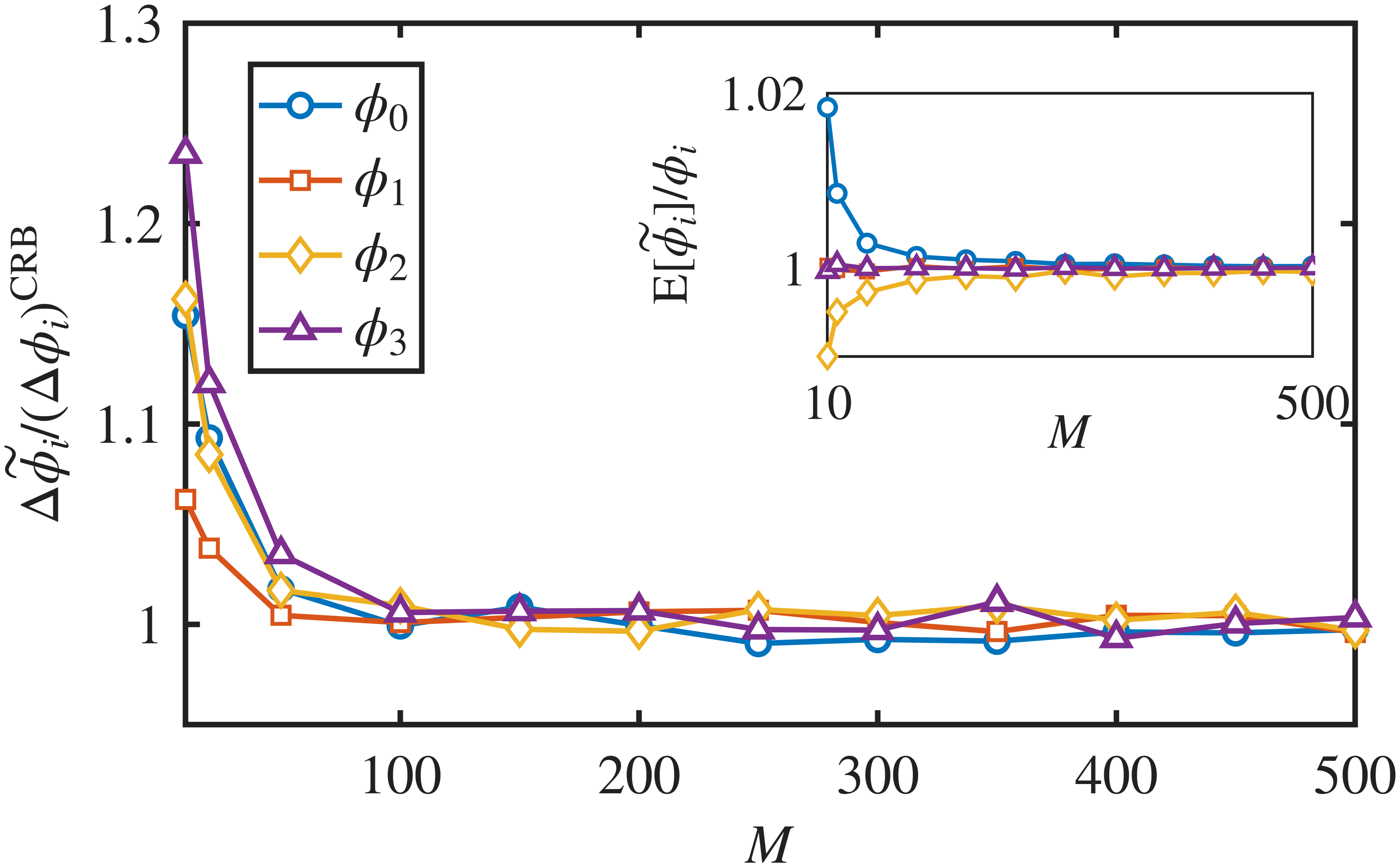}
    \caption{Maximum likelihood estimation of the four parameters under the two-port homodyne detection for $N=10$ and $\beta=1/2$. The main panel shows the normalized CRB $\Delta\widetilde{\phi}_i/(\Delta\phi_i)^{\mathrm{CRB}}$ with $i=0,1,2,3$ as a function of the number of measurement repetitions $M$, where $\Delta\widetilde{\phi}_i$ is the standard deviation of the estimator and $(\Delta\phi_i)^{\mathrm{CRB}}$ is the corresponding marginal CRB in Eqs.~\eqref{eq:CRB_phi0}--\eqref{eq:CRB_omega}. The convergence of all four curves to unity confirms the saturation of the CRBs already for a relatively small number $M$ of experimental iterations, typically $M\approx 100$. In the inset, we plot the ratio between the expected value $\mathrm{E}[\widetilde{\phi}_i]$ of the maximum-likelihood estimator and the true value $\phi_i$. Here, $k_1=k_2=0.5$ and $k_3=0$.}
    \label{fig:CRB_vs_M}
\end{figure}
With the resource scaling \(N_{\mathrm{s}}=\beta N\) and \(N_{\mathrm{c}}=(1-\beta)N\), where $\beta\in(0,1)$, the total FIM can be asymptotically written by combining the two contributions as 
\begin{equation}
F = N^2\,\mathcal{F}(k_1,k_2,k_3,\beta)+{O}(N),
\label{eq:Ftot_asym_main}
\end{equation}
where $\mathcal{F}=\beta^2\mathcal{F}^{\Sigma}+\beta(1-\beta)\mathcal{F}^{\bm{\mu}}$. The \({O}(N^{2})\) term of the Fisher matrix is sufficient to establish Heisenberg-scaling behavior of the parameter sensitivities for the non-singular coefficient matrix $\mathcal{F}$.
For the multiparameter analysis, the results of the calculation for $F^{-1}$ are shown in Fig.~\ref{fig:effec_FIM}, which shows that the $N^2$-rescaled scalar quantity $N^2\,\mathrm{Tr}[F^{-1}]$ approaches an $N$-independent plateau, confirming the asymptotic Heisenberg scaling $\mathrm{Tr}[F^{-1}]\propto1/N^{2}$. In particular, for any (locally) unbiased estimator $\widetilde{\bm{\phi}}=(\widetilde{\phi}_0,\widetilde{\phi}_1,\widetilde{\phi}_2,\widetilde{\phi}_3)$ of $\bm{\phi}$ and for \(M\) independent experimental repetitions, the marginal estimator variance denoted by $\mathrm{Var}(\widetilde{\bm{\phi}})\equiv\Delta^2\widetilde{\bm \phi}$ asymptotically read~\cite{helstrom1969quantum}
 \begin{align}
  \Delta^2\widetilde{\phi}_0
  &\ge(\Delta^2\phi_0){^{\mathrm{CRB}}}=
  \frac{1}{M\, N^{2}}\,
  \bigl[\mathcal{F}^{-1}\bigr]_{11}, \label{eq:CRB_phi0}\\
 \Delta^2\widetilde{\phi}_1
  &\ge(\Delta^2\phi_1)^{\mathrm{CRB}}=
  \frac{1}{M\, N^{2}}\,
  \bigl[\mathcal{F}^{-1}\bigr]_{22}, \label{eq:CRB_phi}\\
  \Delta^2\widetilde{\phi}_2
  &\ge(\Delta^2\phi_2)^{\mathrm{CRB}}=
  \frac{1}{M\, N^{2}}\,
  \bigl[\mathcal{F}^{-1}\bigr]_{33}, \label{eq:CRB_psi}\\
   \Delta^2\widetilde{\phi}_3
  &\ge(\Delta^2\phi_3)^{\mathrm{CRB}}=
  \frac{1}{M\, N^{2}}\,
  \bigl[\mathcal{F}^{-1}\bigr]_{44}. \label{eq:CRB_omega}
\end{align}
These bounds demonstrate simultaneous Heisenberg-scaling estimation for all four parameters, provided the \(k_i\) values are chosen such that the leading-order coefficient matrix $\mathcal{F}$ is nonsingular. The coefficient matrix $\mathcal{F}(k_1,k_2,k_3,\beta)$ in Eq.~\eqref{eq:Ftot_asym_main} is generically nonsingular and loses rank only at the singular choices $k_1=-k_2$ or $k_3^2=k_1k_2$~\cite{Supp}. Figure~\ref{fig:effec_FIM} shows that different admissible choices of \(k_1,k_2,k_3\) preserve the \(1/N^2\) scaling, while changing only the prefactors. For example, for $\beta=1/2$, $k_1=k_2=1/2$ , and $k_3=0$, one finds $ \mathrm{diag}(\mathcal{F}^{-1})=(1,4,4,1)$, so the corresponding CRBs therefore scale as $\Delta^2\widetilde{\phi}_0 \ge1/(M N^2),\,\Delta^2\widetilde{\phi}_1\ge4/(M N^2),\,\Delta^2\widetilde{\phi}_2\ge4/(M N^2),$ and $\Delta^2\widetilde{\phi}_3\ge1/(M N^2)$ at leading order. The balanced resource split \(\beta=1/2\) (i.e., $N_s=N_c$) maximizes the leading contribution \(F^{\bm{\mu}}\) in Eq.~\eqref{eq:AsymFmu}, since it scales as $N_s\,N_c$. By contrast, minimizing the scalar bound (sum of the CRBs) \(\mathrm{Tr}[F^{-1}]\) typically favors \(\beta>1/2\), since three of the four parameters derive their leading information from the squeezing-dominated contribution \(F^\Sigma\propto N_s^2\). For simplicity, we choose $\beta=1/2$ throughout the text. 

\begin{figure}
    \centering
    \includegraphics[width=\linewidth]{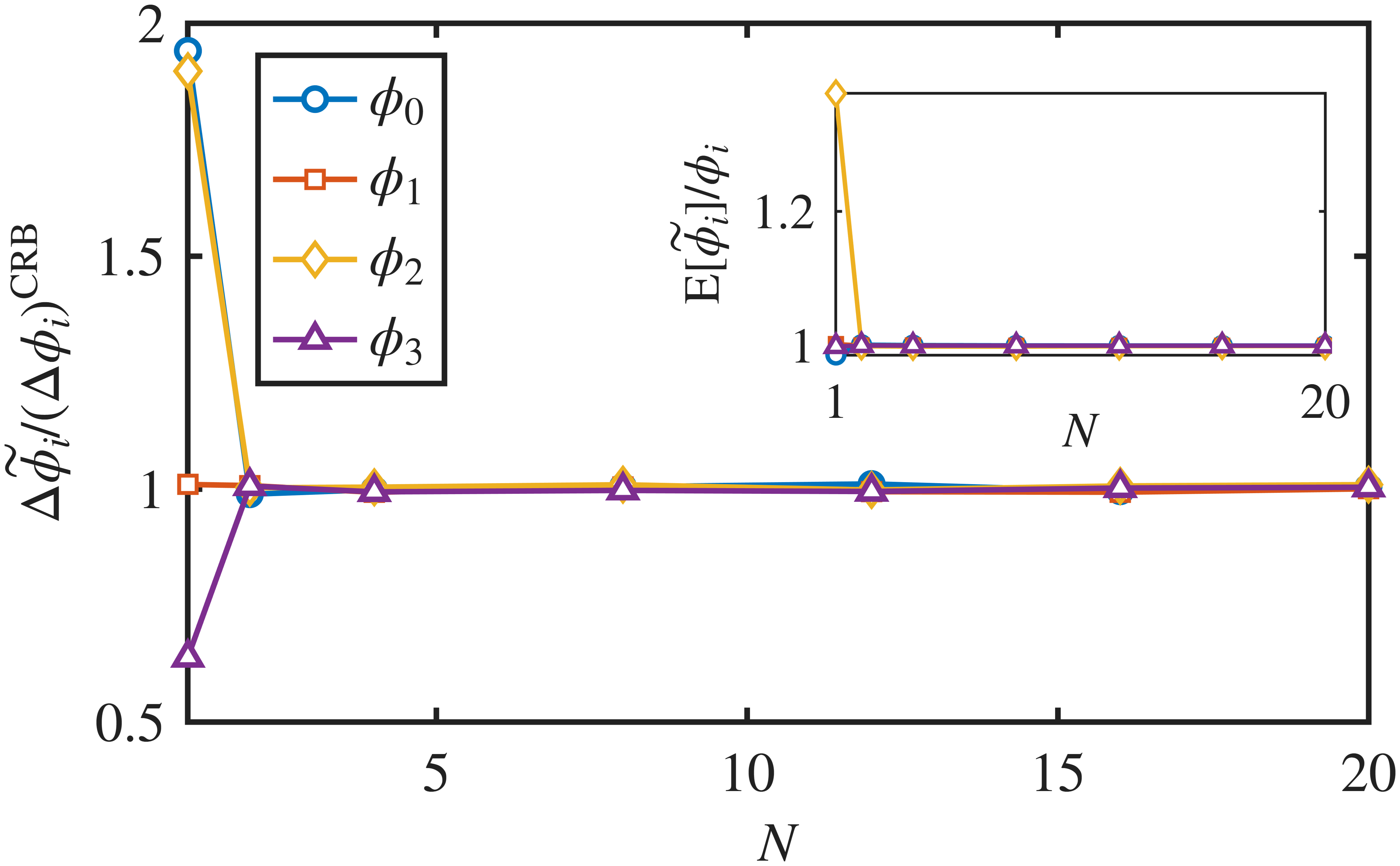}
    \caption{Maximum likelihood estimation of $\bm{\phi}$ as a function of the total mean photon number $N$ at a fixed sample size $M=200$. The main panel shows convergence of the normalized CRB $\Delta\widetilde{\phi}_i/(\Delta\phi_i)^{\mathrm{CRB}}$ for each parameter in Eqs.~\eqref{eq:CRB_phi0}--\eqref{eq:CRB_omega} saturating the CRBs already at low photon numbers. The inset shows the unbiasedness ratio $\mathrm{E}[\widetilde{\phi}_i]/\phi_i$, confirming the estimator remains effectively unbiased for small values of $N$. Here, $k_1=k_2=0.5,k_3=0$ and $\beta=1/2$ for all $N$.}
    \label{fig:CRB_vs_N}
\end{figure}
Further, we go beyond stating the bounds and demonstrate the attainability of the CRBs by applying maximum-likelihood estimation (MLE). In particular, the estimators are obtained from the outcomes of $M$ sampling measurements from the probability distribution at the interferometer output in Eq.~\eqref{eq:Gaussian_pdf_main}. We show that the errors in the estimators approach the CRB associated with the FIM obtained in the asymptotic limit already for $M$ of order 100 sampling measurements. We implement the MLE for all four parameters and show that it asymptotically saturates the CRBs in Eqs.~\eqref{eq:CRB_phi0}--\eqref{eq:CRB_omega}, thus achieving Heisenberg-scaling precision. The derivation of the maximum-likelihood estimator for this scheme is given in SM~\cite{Supp}. Figures~\ref{fig:CRB_vs_M} and~\ref{fig:CRB_vs_N} illustrate its behavior, confirming the saturation of the CRB for all parameters simultaneously already for a sample size of the order $M\approx 100$ and for a relatively small average number of photons in the squeezed and coherent states. The insets show the ratio of the expected value $\mathrm{E}[\widetilde{\bm{\phi}}]$ of the estimator and the true value $\bm{\phi}$ approaching unity, confirming that the estimator bias vanishes for the same order of sample size $M$ and the total mean photon number $N$. These results provide direct numerical validation that the multiparameter MLEs saturate the CRBs and achieve the Heisenberg-scaling precision. Our scheme remains robust for the different parameter values $\phi_0,\phi_1$, and $\phi_2$ varying over their periodic domains $[0,2\pi)$ upon tuning the LO phases~\eqref{Eq:LOtuning}  (see Fig.~\ref{fig:CRB_Bias_M_combined} in SM~\cite{Supp}), as far as $\phi_3$ is operated close to $\pi/4$. 

\paragraph*{Robustness analysis---}
For imperfect homodyne detectors with efficiency $\eta$, the Heisenberg scaling for all four parameters remains observable in the finite-photon regime $N\ll N_\eta=2\eta/(1-\eta)$. A closed-form analysis of the homodyne detection inefficiencies is given in Sec.~\ref{Appendix D} of the SM.
At the high-efficiency detection $\eta=0.99$, i.e., $1\%$ detection inefficiency at each homodyne detector and for the low-photon-number regime $N=5$--$10$, the uncertainties given by CRBs increase only by about \(2.5\%\)--\(5.1\%\) for the parameters $\phi_0,\phi_2,\phi_3$ which are dominantly sensitive to the fluctuation of the homodyne signal, and by about \(1.3\%\)--\(2.5\%\) for $\phi_1$ which is determined by the signal mean of the homodyne.

\paragraph*{Conclusions---}
We have introduced a fully Gaussian and experimentally realistic scheme for the simultaneous estimation of the full set of four real parameters defining a two-channel linear-optical transformation $\mathrm{U}(2)$ using a displaced two-mode squeezed probe and two-port balanced homodyne detection. Working within the classical estimation framework defined by the homodyne statistics, we derive the FIM and corresponding multiparameter CRBs, and we identify operating conditions under which the attainable precision exhibits Heisenberg scaling.

Our results show that three of the parameters are predominantly encoded in the squeezing controlled fluctuations, so their precision is governed by the covariance contribution $F^{\Sigma}$, whereas, the parameter $\phi_1$ is accessed primarily through the signal of the homodyne outcome, and its Fisher information is therefore given by $F^{\bm{\mu}}$, highlighting the essential role of the displacement to estimate all four parameters simultaneously.
Our analysis also shows the practical conditions required to achieve the Heisenberg scaling. The mode-mixing angle $\phi_3$ must be operated close to the balanced point $\pi/4$, in addition, optimal sensitivity requires tuning of the LO phases $\theta_1$ and $\theta_2$ using coarse prior estimates of $\phi_0$ and $\phi_2$. 

Finally, we demonstrate attainability using the exact Gaussian likelihood for the homodyne outcomes. We construct a maximum-likelihood estimator and show that it approaches the CRB already for modest data size, of the order of 100 experimental runs, and even at a few mean photon numbers. The saturation of the CRBs for fewer experimental iterations and photon numbers is essential for realistic sensing and device characterization scenarios. Importantly, our scheme remains robust against small homodyne detector inefficiencies in this low-photon regime. Extending the computation of sensitivity bounds to include the impact of optical noise and losses at the different stages of the sensor network is an important direction for future work.

\paragraph*{Acknowledgments---} This work was partially supported by Xairos System Inc. VT also acknowledges partial support from the Air Force Office of Scientific Research under award number FA8655-23-17046. PF was partially supported by Istituto Nazionale di Fisica Nucleare (INFN) through the project ``QUANTUM'', by the Italian National Group of Mathematical Physics (GNFM-INdAM), and by the Italian funding within the ``Budget MUR - Dipartimenti di Eccellenza 2023--2027'' - Quantum Sensing and Modelling for One-Health (QuaSiModO). We acknowledge useful discussions with Danilo Triggiani.

\paragraph*{Data Availability---}The data that support the findings of this article are publicly available~\cite{rai_2026_19132674}.
\twocolumngrid 
\bibliography{mybib} 

\clearpage
\onecolumngrid
\setcounter{section}{0}
\setcounter{equation}{0}
\setcounter{figure}{0}
\setcounter{table}{0}
\renewcommand{\thesection}{\Roman{section}}
\renewcommand{\theequation}{S\arabic{equation}}
\renewcommand{\thefigure}{S\arabic{figure}}
\renewcommand{\thetable}{S\arabic{table}}

\makeatletter
\renewcommand{\theHsection}{S\Roman{section}}
\renewcommand{\theHequation}{S\arabic{equation}}
\renewcommand{\theHfigure}{S\arabic{figure}}
\renewcommand{\theHtable}{S\arabic{table}}
\makeatother
\begin{center}
{\large\bf Supplemental Material}

\vspace{0.4cm}

{\bf Heisenberg-scaling characterization of a two-channel optical network via two-port homodyne detection}

\vspace{0.4cm}
Atmadev Rai,$^{1,2}$ Paolo Facchi,$^{3,4}$ and Vincenzo Tamma$^{1,2,5}$

\vspace{0.2cm}

{\small
$^{1}$School of Mathematics and Physics, University of Portsmouth, Portsmouth PO1 3QL, United Kingdom\\
$^{2}$Quantum Science and Technology Hub, University of Portsmouth, Portsmouth PO1 3QL, United Kingdom\\
$^{3}$Dipartimento di Fisica, Universit\`{a} di Bari \textup{\&} Politecnico di Bari, I-70126 Bari, Italy\\
$^{4}$INFN, Sezione di Bari, I-70126 Bari, Italy\\
$^{5}$Institute of Cosmology and Gravitation, University of Portsmouth, Portsmouth PO1 3FX, United Kingdom
}
\end{center}
\vspace{0.5cm}
%\maketitle
%\onecolumngrid
%\renewcommand{\theequation}{S\arabic{equation}}  
\section{Gaussian statistics: Two-port homodyne detection} \label{Appendix A} 
  \setcounter{equation}{0} 
In this section, we derive the Gaussian probability distribution~\eqref{eq:Gaussian_pdf_main} for two-port homodyne detection at the output of a two-channel unitary $\mathrm{U}(2)$ network probed by a TMSS,
\begin{equation}
\ket{\psi_{\mathrm{in}}}
= \hat{D}_1(\alpha_1)\,\hat{D}_2(\alpha_2)\,\hat{S}_{12}(\xi)\,\ket{0,0},
\label{SM:INPUTstate}
\end{equation}
where $\hat{D}_i(\alpha_i)$ is the single-mode displacement operator acting on mode $i$ with
$\alpha_i = |\alpha_i|e^{i\beta_i}$ for $i = 1,2$, and $\hat{S}_{12}(\xi)$ is the two-mode squeezing operator with complex squeezing parameter $\xi = r e^{i\theta}$. For simplicity, and without loss of optimality for the quantum Fisher information matrix, we restrict to equal real displacements in both input modes,
\(|\alpha_1| = |\alpha_2| = |\alpha|\) and \(\beta_1 = \beta_2 = 0\), and to real squeezing, $\xi = r$ with $\theta = 0$, which satisfies the optimality condition $\theta - \beta_1 - \beta_2 = 0$~\cite{537s-r3t8}.

Adopting the quadrature ordering
\(\mathbf{z} = (x_1,\,p_1,\,x_2,\,p_2)^{\mathsf T}\),
the initial displacement vector of the input state~\eqref{SM:INPUTstate} is $\bm d_0=\sqrt{2}\,|\alpha|(1,0,1,0)^\mathsf{T},$ and the covariance matrix of the two-mode squeezed vacuum $\hat{S}_{12}(r)\ket{0,0}$ is
\begin{equation}
\Gamma_0
=
\frac{1}{2}\begin{pmatrix}
\cosh(2r) & 0 & \sinh(2r) & 0 \\
0 & \cosh(2r) & 0 & -\sinh(2r) \\
\sinh(2r) & 0 & \cosh(2r) & 0 \\
0 & -\sinh(2r) & 0 & \cosh(2r)
\end{pmatrix}.
\end{equation}
After the action of the two-channel network $U_{\bm{\phi}}$, the displacement and covariance matrix transform as
\begin{equation}
 \bm d_{\bm{\phi}} = R_{\bm{\phi}}\,\bm{d_0},
\qquad
 \Gamma_{\bm{\phi}} = R_{\bm{\phi}}\,\Gamma_0\,R_{\bm{\phi}}^{\mathsf T},
\end{equation}
where $R_{\bm{\phi}}$ is the real orthogonal symplectic matrix associated with the unitary $U_{\bm{\phi}}$,
\begin{equation}
    R_{\bm{\phi}} =
    \begin{pmatrix}
        \Re[U_{\bm{\phi}}] & -\Im[U_{\bm{\phi}}] \\
        \Im[U_{\bm{\phi}}] & \Re[U_{\bm{\phi}}]
    \end{pmatrix}.
 \label{App:R}
\end{equation}

We then perform homodyne measurements on both output ports of the network. On mode \(j = 1,2\) we measure the quadrature $\hat x_{\theta_j}^{(j)}=\frac{1}{\sqrt{2}}\bigl(e^{-i\theta_j}\hat a_{j,\text{out}} + e^{i\theta_j}\hat a_{j,\text{out}}^\dagger\bigr)$, rotated by the LO phase \(\theta_j\). Collecting the two measurement outcomes into the vector \(\bm{x} = (x_1, x_2)^\mathsf T\), where $x_j$ is the eigenvalue of $\hat x^{(j)}_{\theta_j}$, the measured observables can be written as linear combinations of the output quadratures,
\begin{equation}
    \hat{\bm{x}} = M_{\hom} \,\hat{\mathbf{z}}_{\bm{\phi}},
\end{equation}
where \(\hat{\mathbf{z}}_{\bm{\phi}}\) is the output quadrature operator vector and the \(2\times 4\) matrix $M_{\hom}$ selects the rotated quadratures measured by balanced homodyne detection:
\begin{equation}
    M_{\hom}
    =
    \begin{pmatrix}
        \cos(\theta_1) & \sin(\theta_1) & 0 & 0 \\
        0 & 0 & \cos(\theta_2) & \sin(\theta_2)
    \end{pmatrix}.
\end{equation}

Since the input state is Gaussian and the transformation $U_{\bm{\phi}}$ is linear and passive, the joint statistics of $(x_1,x_2)$ remain Gaussian. The corresponding mean vector and covariance matrix are
   $\bm{\mu}_{\bm{\phi}} = M_{\hom}\,\bm d_{\bm{\phi}}$ and $\Sigma_{\bm{\phi}}
    =
    M_{\hom}\,\Gamma_{\bm{\phi}}\,M_{\hom}^{\mathsf T}$, respectively. The joint probability density of observing outcomes \(\bm{x} = (x_1,x_2)^\mathsf T\) is therefore
\begin{equation}
   p(\bm{x}\mid\bm{\phi})
    =
    \frac{1}{2\pi\sqrt{\det \Sigma_{\bm{\phi}}}}
    \exp\!\left[
    -\frac{1}{2}
    \big(\bm{x} - \bm{\mu}_{\bm{\phi}}\big)^{\mathsf T}
    \Sigma^{-1}_{\bm{\phi}}
    \big(\bm{x} - \bm{\mu}_{\bm{\phi}}\big)
    \right],
    \label{EqA:probdistribution}
\end{equation}
as in Eq.~\eqref{eq:Gaussian_pdf_main} in the main text. For the parameter vector
\(\bm{\phi} = (\phi_0,\phi_1,\phi_2,\phi_3)\)
of a two-channel network \(U_{\bm{\phi}}\), the mean vector and covariance matrix can be written explicitly as
\begin{equation}
    \bm{\mu}_{\bm{\phi}}
    =
    \sqrt{2}\,|\alpha|
    \begin{pmatrix}
\cos(\phi_3)\,
\cos\left(\tfrac{1}{2}(2\theta_1 - \phi_1 - 2\phi_0 - \phi_2)\right)
+
\sin(\phi_3)\,
\cos\left(\tfrac{1}{2}(2\theta_1 + \phi_1 - 2\phi_0 - \phi_2)\right)
\\
\cos(\phi_3)\,
\cos\left(\tfrac{1}{2}(2\theta_2 + \phi_1 - 2\phi_0 + \phi_2)\right)
-
\sin(\phi_3)\,
\cos\left(\tfrac{1}{2}(2\theta_2 - \phi_1 - 2\phi_0 + \phi_2)\right)
    \end{pmatrix},
    \label{EqA:mu}
\end{equation}
\begin{equation}
    \Sigma_{\bm{\phi}}
    =
    \frac{1}{2}
    \begin{pmatrix}
        \cosh(2r)
        - \sin(2\phi_3)\,
          \cos\!\,\bigl(2\theta_1 - 2\phi_0 - \phi_2\bigr)\,
          \sinh(2r)
        &
       -\cos(2\phi_3)\,
        \sinh(2r)\,
        \cos\!\,\bigl(\theta_1 + \theta_2 - 2\phi_0\bigr)
        \\
       -\cos(2\phi_3)\,
        \sinh(2r)\,
        \cos\!\,\bigl(\theta_1 + \theta_2 - 2\phi_0\bigr)
        &
        \cosh(2r)
        + \sin(2\phi_3)\,
          \cos\!\,\bigl(2\theta_2 - 2\phi_0 + \phi_2\bigr)\,
          \sinh(2r)
    \end{pmatrix}.
    \label{EqA:Sigma}
\end{equation}

\section{Derivation of the Fisher information matrix} \label{Appendix B}     
\subsection{Asymptotic form of  \texorpdfstring{$F^{\Sigma}$}{F-Sigma} in Eq.~\eqref{eq:AsymFSigma}}
We begin with the first term in Eq.~\eqref{eq:FIMGaussian_main} of the main text,
\begin{equation}
F^{\Sigma}_{ij}=\frac{1}{2}\,
  \mathrm{Tr}\!\left[
    \Sigma_{\bm{\phi}}^{-1}
    \big(\partial_i \Sigma_{\bm{\phi}}\big)
    \Sigma_{\bm{\phi}}^{-1}
    \big(\partial_j \Sigma_{\bm{\phi}}\big)
  \right],
    \label{App:FIMSigma}  
\end{equation}
where \(\Sigma_{\bm{\phi}}\) is given explicitly in Eq.~\eqref{EqA:Sigma}. Expressed in terms of the average photon number of the TMSS, $N_s=2\sinh^2(r)$, the covariance matrix reads
\begin{equation}
\Sigma_{\bm{\phi}}=\frac{1}{2}
\begin{pmatrix}
N_s+1 -\sqrt{N_s(N_s+2)}\,\cos(2\theta_1-2\phi_0-\phi_2)\,\sin (2\phi_3)
&
-\sqrt{N_s(N_s+2)}\cos(\theta_1+\theta_2-2\phi_0)\,\cos (2\phi_3)\\
-\sqrt{N_s(N_s+2)}\cos(\theta_1+\theta_2-2\phi_0)\,\cos (2\phi_3)
&
N_s+1+\sqrt{N_s(N_s+2)}\,\cos(2\theta_2-2\phi_0+\phi_2)\,\sin (2\phi_3)
\end{pmatrix}.
\label{eq:B_Sigma_explicit}
\end{equation}
Since $\sqrt{N_s(N_s+1)}={O}(N_s)$ and all trigonometric factors are bounded, each entry of $\Sigma_{\bm{\phi}}$ is at most ${O}(N_s)$, and same holds for its parameter derivatives, i.e. $\partial_{\phi_i}\Sigma_{\bm{\phi}}={O}(N_s)$. Therefore for generic LO phases, the determinant scales as $\mathrm{det}(\Sigma_{\bm{\phi}})={O}(N_s^2)$, while the adjoint matrix $\mathrm{adj}(\Sigma_{\bm{\phi}})$ has entries of order ${O}(N_s)$. Using $\Sigma_{\bm{\phi}}^{-1}=\mathrm{adj}(\Sigma_{\bm{\phi}})/\mathrm{det}(\Sigma_{\bm{\phi}})$ therefore yields $\Sigma_{\bm{\phi}}^{-1}={O}(N_s^{-1})$. Consequently, the matrix product inside the trace in Eq.~\eqref{App:FIMSigma} is ${O}(N_s^0)$, resulting in $F^{\Sigma}_{ij}={O}(N_s^0)$.

Therefore, to achieve the Heisenberg-scaling sensitivity, we tune the homodyne phases to the minimum-variance quadratures according to condition~\eqref{Eq:LOtuning} of the main text. We tune the LO phases as $\theta_1=f_1+k_1/N_s$ and $\theta_2=f_2+k_2/N_s$, where $f_1=\phi_0+\phi_2/2$ $f_2=\pi/2+\phi_0-\phi_2/2$ are the quadrature angles that minimize the variances of modes $1$ and $2$, respectively. In practice, these angles can be identified from a coarse estimation of the output second moments and by locating the corresponding minima. We also operate close to a balanced beam splitter, $\phi_3=\pi/4+\delta\phi_3$, and in the asymptotic expansion we take $\delta\phi_3=k_3/N_s$, where $k_1,k_2, k_3$ are $N$-independent constants. Under these conditions, the covariance matrix admits the asymptotic form
\begin{equation}
    \Sigma_{\bm{\phi}}=\begin{pmatrix}
    \frac{1+4k_1^2+4k_3^2}{4 N_s}+{O}(\frac{1}{N_s^2})&-\frac{(k_1+k_2)\,k_3}{N_s}+{O}(\frac{1}{N_s^2})\\-\frac{(k_1+k_2)\,k_3}{N_s}+{O}(\frac{1}{N_s^2})&  \frac{1+4k_2^2+4k_3^2}{4 N_s}+{O}(\frac{1}{N_s^2})
    \end{pmatrix}.
    \label{App:AsymSigma}
\end{equation}
Evaluating $\Sigma_{\bm{\phi}}^{-1}$ and $\partial_i\Sigma_{\bm{\phi}}$ in the same conditions and substituting into Eq.~\eqref{eq:FIMGaussian_main}, the FIM $F^{\Sigma}$ asymptotically reads as in Eq.~\eqref{eq:AsymFSigma} in the main text
\begin{equation}
F^{\Sigma}
=
N_s^{2}\,\mathcal{F}^\Sigma+{O}(N_s),
\label{eq:FSigma_compact_new}
\end{equation}
with the coefficient matrix
\begin{equation}
\mathcal{F}^{\Sigma}=
\frac{1}{\Lambda^{2}}\,
\begin{pmatrix}
D_{1} & 0   & A   & B   \\
0     & 0   & 0   & 0   \\
A     & 0   & D_{3} & C \\
B     & 0   & C   & D_{4}
\end{pmatrix},
\label{eq:FSigma_coiffiecnt}
\end{equation}
where $\Lambda \equiv (1+4k_1^{2})(1+4k_2^{2})+8(1-4k_1k_2)k_3^{2}+16k_3^{4}$, and the nonzero matrix elements are
\begin{align}
D_{1}
&=
32\Bigl[
k_2^{2}
+k_1^{2}\bigl(1+16k_2^{2}(1+k_1^{2}+k_2^{2})\bigr)
+2k_3^{2}
-32k_1k_2(1+k_1^{2}-k_1k_2+k_2^{2})k_3^{2}
\nonumber\\
&\hspace{2.2cm}
+16(1+k_1^{2}-4k_1k_2+k_2^{2})k_3^{4}
+32k_3^{6}
\Bigr],
\label{eq:D1_def_new}
\\
D_{3}
&=
8\Bigl[
k_2^{2}
+k_1^{2}\bigl(1+16k_2^{2}(1+k_1^{2}+k_2^{2})\bigr)
-8(-1+4k_1k_2)(k_1^{2}+k_2^{2})k_3^{2}
+16(k_1^{2}+k_2^{2})k_3^{4}
\Bigr],
\label{eq:D3_def_new}
\\
D_{4}
&=
16\Bigl[
(1+4k_1^{2})(k_1+k_2)^{2}(1+4k_2^{2})
-4\Bigl(-1+2\bigl(k_1^{2}+6k_1k_2+4k_1^{3}k_2+k_2^{2}+4k_1k_2^{3}\bigr)\Bigr)k_3^{2}
\nonumber\\
&\hspace{2.2cm}
+16(2+k_1^{2}-6k_1k_2+k_2^{2})k_3^{4}
+64k_3^{6}
\Bigr].
\label{eq:D4_def_new}
\\
A
&=
-16\,(k_1^2-k_2^2)\,
\bigl(-1+4k_1k_2-4k_3^{2}\bigr)\,
\bigl(1+4k_1k_2-4k_3^{2}\bigr),
\label{eq:A_def_new}
\\[2pt]
B
&=
64\,(k_1+k_2)k_3\,
\bigl(-1+4k_1k_2-4k_3^{2}\bigr)\,
\bigl(1+4k_1k_2-4k_3^{2}\bigr),
\label{eq:B_def_new}
\\[2pt]
C
&=
-16\,(k_1-k_2)k_3\,
\bigl(-1-2k_2+k_1(-2+4k_2)-4k_3^{2}\bigr)\,
\bigl(-1+2k_2+k_1(2+4k_2)-4k_3^{2}\bigr).
\label{eq:C_def_new}
\end{align}
Equations~\eqref{eq:D1_def_new}–\eqref{eq:C_def_new} collect all
\({O}(N_s^{2})\) coefficients, which are the only ones that contribute to Heisenberg scaling.
\subsection{Asymptotic form of \texorpdfstring{$F^{\bm{\mu}}$}{F-bar-mu} in Eq.~\eqref{eq:AsymFmu}}
The FIM contribution from the derivatives of the mean homodyne signal is given by
\begin{equation}
F^{\bm{\mu}}_{ij} =(\partial_i\bm{\mu}_{\bm{\phi}})^{\mathsf{T}}\,\Sigma_{\bm{\phi}}^{-1}\,(\partial_j\bm{\mu}_{\bm{\phi}}),    
\end{equation} 
and depends on both the coherent displacement and the squeezing through \(\bm{\mu}_{\bm{\phi}}\) and \(\Sigma_{\bm{\phi}}\). In the same conditions~\eqref{Eq:LOtuning} and for \(\phi_3=\pi/4+k_3/N_s\) by substituting the parameter derivatives of the mean $\bm{\mu}_{\bm{\phi}}$~\eqref{EqA:mu} and $\Sigma_{\bm{\phi}}^{-1}$ into $F^{\bm{\mu}}$ yields, to the leading order,
\begin{equation}
F^{\bm{\mu}}
=
N_sN_c\,\mathcal{F}^{\bm{\mu}}(k_1,k_2,k_3,\phi_1)
\;+\;{O}(N)
\label{eq:FmuAsym_compact}
\end{equation}
where the coefficient matrix is independent of \(N\). Explicitly,
\begin{equation}
\mathcal{F}^{\bm{\mu}}
=
\frac{2}{\Lambda}\,
\begin{pmatrix}
0&0&0&0\\
0&D_2&0&0\\
0&0&0&0\\
0&0&0&0
\end{pmatrix},
\label{eq:Fmu2_matrix}
\end{equation}
with
\begin{equation}
D_2=1+2k_1^{2}+2k_2^{2}+4k_3^{2}+2(k_1+k_2)\big[(k_1-k_2)\cos(\phi_1)+2k_3\sin(\phi_1)\big].
\end{equation}
We find that the only entry with a leading
\({O}(N_sN_c)={O}(N^2)\) contribution is the $[F^{\bm{\mu}}]_{22}$ component, while all other entries are at most \({O}(N)\) or \({O}(N^0)\), since \(N_{s,c}={O}(N)\).
\subsection{Total asymptotic FIM and CRBs}
\begin{figure}
    \centering
    \includegraphics[width=\linewidth]{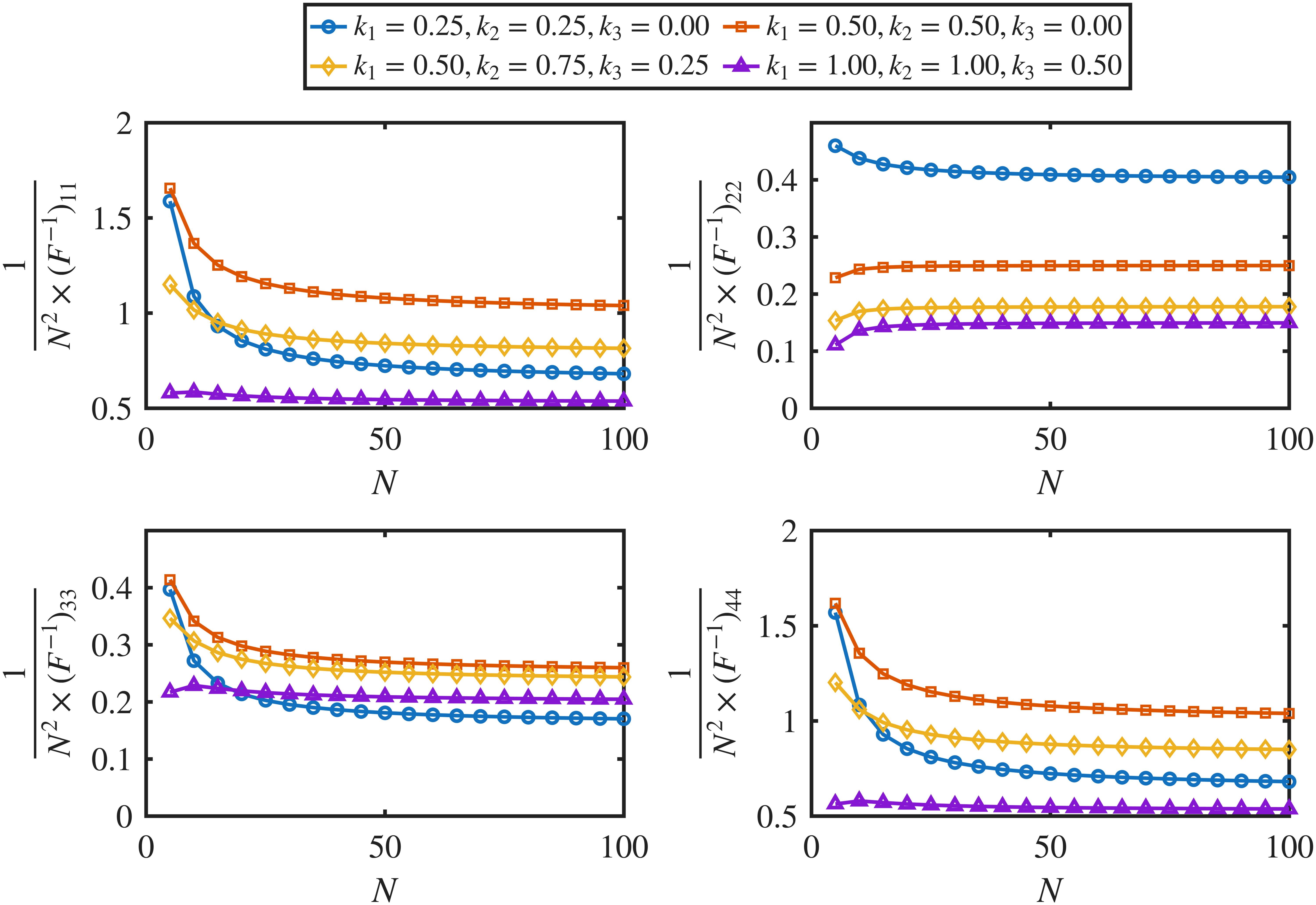}
    \caption{Each panel shows the normalized effective Fisher information $1/(N^2\,(F^{-1})_{ii})$ plotted versus the total mean photon number $N$ for each of the four estimated parameters $\phi_0,\phi_1,\phi_2$, and $\phi_3$, respectively. Curves compare different LO detuning constants $k_1, k_2$ and balanced beam-splitter offset $k_3$. A plateau of $1/(N^2\,(F^{-1})_{ii})$ at large $N$ indicates that asymptotically $(F^{-1})_{ii}\propto 1/N^2$, i.e., a Heisenberg scaling of the marginal variance for that parameter. The vertical separation between curves reflects the prefactor of the asymptotic scaling: higher plateaus correspond to larger effective Fisher information and, therefore, higher achievable precision. Differences between panels highlight that the optimal working point need not be identical for all parameters. However, the asymptotic Heisenberg scaling behavior holds for different values of $k_1,k_2,k_3$.}
    \label{fig:Eff_FIM_diagonal}
\end{figure}
Introducing the resource split $N_s=\beta N$, $N_c=(1-\beta)N$ for any value $0<\beta<1$ independent of $N$, and combining the two contributions of the FIM in Eq.~\eqref{eq:FSigma_compact_new} and Eq.~\eqref{eq:FmuAsym_compact}, the total FIM up to the leading order reads,
\begin{equation}
    F = F^{\Sigma} + F^{\bm{\mu}} = N^{2}\,\mathcal{F}(k_1,k_2,k_3,\beta) + {O}(N),
    \label{eq:FtotAsym}
\end{equation}
where the leading-order coefficient matrix $\mathcal{F}=\mathcal{F}^{\Sigma}+\mathcal{F}^{\bm{\mu}}$ is
\begin{equation}
    \mathcal{F}
=
  \frac{\beta^{2}}{\Lambda^{2}}\,
  \begin{pmatrix}
    D_{1} & 0   & A   & B   \\
    0     & 0   & 0   & 0   \\
    A     & 0   & D_{3} & C \\
    B     & 0   & C   & D_{4}
  \end{pmatrix}
    \;+\;
\frac{2(1-\beta)\beta}{\Lambda}
    \begin{pmatrix}
        0 & 0 & 0 & 0 \\
        0 & D_2 & 0 & 0 \\
        0 & 0 & 0 & 0 \\
        0 & 0 & 0 & 0
    \end{pmatrix}.
    \label{eq:F2_def}
\end{equation}
Equation~\eqref{eq:F2_def} shows how the
Heisenberg scaling of the total Fisher information depends on the allocation of photons between squeezing and displacement through
\(N_{\mathrm{s}}\) and \(N_{\mathrm{c}}\). The multiparameter Cram\'{e}r–Rao bound for any unbiased estimator
\(\widetilde{\bm{\phi}}\) is given by $ \mathrm{Cov}\bigl[\widetilde{\bm{\phi}}\bigr]
    \geq
F^{-1}/M$. 
Therefore, using Eq.~\eqref{eq:FtotAsym}, the marginal variances of the four parameters $\phi_0,\phi_1,\phi_2,\phi_3$ satisfy, at leading order,
\begin{align}
    \Delta^2 \widetilde{\phi}_0
    &\;\ge\;
    \frac{1}{M\,N^2}\,\bigl[\mathcal{F}^{-1}\bigr]_{11}+O\bigg(\frac{1}{MN^3}\bigg),\label{Appeq:CRB_diagonal_phi0}\\
    \Delta^2 \widetilde{\phi}_1
    &\;\ge\;
      \frac{1}{M\,N^2}\,\bigl[\mathcal{F}^{-1}\bigr]_{22}+O\bigg(\frac{1}{MN^3}\bigg),\label{Appeq:CRB_diagonal_phi}\\
    \Delta^2 \widetilde{\phi}_2
    &\;\ge\;
      \frac{1}{M\,N^2}\,\bigl[\mathcal{F}^{-1}\bigr]_{33}+O\bigg(\frac{1}{MN^3}\bigg),\label{Appeq:CRB_diagonal_psi}\\
    \Delta^2 \widetilde{\phi}_3
    &\;\ge\;
     \frac{1}{M\,N^2}\,\bigl[\mathcal{F}^{-1}\bigr]_{44}+O\bigg(\frac{1}{MN^3}\bigg),
\label{Appeq:CRB_diagonal_omega}
\end{align}
as in Eqs.~\eqref{eq:CRB_phi0}--\eqref{eq:CRB_omega} of the main text. Equations~\eqref{Appeq:CRB_diagonal_phi0}–\eqref{Appeq:CRB_diagonal_omega} make explicit that all four parameters exhibit Heisenberg scaling
\(\Delta^2 \widetilde{\phi}_i \propto 1/N^2\), with prefactors determined by the diagonal elements of the coefficient matrix \(\mathcal{F}^{-1}\) (see Fig.~\ref{fig:Eff_FIM_diagonal}).
Since $\mathcal{F}$ is block diagonal, its determinant factorizes as
\begin{equation}
\det (\mathcal{F})
=
\left(\frac{2(1-\beta)\beta}{\Lambda}D_2\right)
\det\!\left(\frac{\beta^{2}}{\Lambda^{2}}
\begin{pmatrix}
D_{1} & A & B\\
A & D_{3} & C\\
B & C & D_{4}
\end{pmatrix}\right),
\end{equation}
since $D_2>0$ for all $\phi_1$ and real $k_1,k_2,k_3$, and for $0<\beta<1$, $\mathcal{F}$ is singular iff the $(\phi_0,\phi_2,\phi_3)$ sub-block is singular. For the coefficients in Eq.~\eqref{eq:F2_def} one finds
\begin{equation}
\det
\begin{pmatrix}
D_{1} & A & B\\
A & D_{3} & C\\
B & C & D_{4}
\end{pmatrix}=
16384\,(k_1+k_2)^2\,(k_1k_2-k_3^2)^2\,\Lambda^3,
\end{equation}
since $\Lambda>0$ for real $k_1,k_2,k_3$. Thus, the leading-order coefficient matrix $\mathcal{F}$ is singular only for 
\begin{equation}
k_1=-k_2, \quad  \text{or} \quad k_3^2=k_1k_2.   
\label{eq:Nonsingular_k}
\end{equation}
At these points, the leading ${O}(N^2)$ Fisher information loses rank, so the four-parameter model is locally unidentifiable at Heisenberg scaling, and at least one parameter combination can only be resolved from subleading terms.
\section{Maximum-Likelihood Estimation}\label{Appendix C}
Here, we derive the maximum-likelihood estimators (MLEs) for the Gaussian statistics generated by two-port homodyne detection. Repeating the experiment \(M\) times yields an i.i.d.\ data set \(\bm{x}_1,\ldots,\bm{x}_M\), where each \(\bm{x}_m=(x_{1,m},x_{2,m})^{\mathsf T}\) is distributed according to \(p(\bm{x}\,|\,\bm{\phi})\) in Eq.~\eqref{eq:Gaussian_pdf_main}, with mean \(\bm{\mu}_{\bm{\phi}}\) and covariance \(\Sigma_{\bm{\phi}}\).
\begin{figure}
    \centering
    \includegraphics[width=\linewidth]{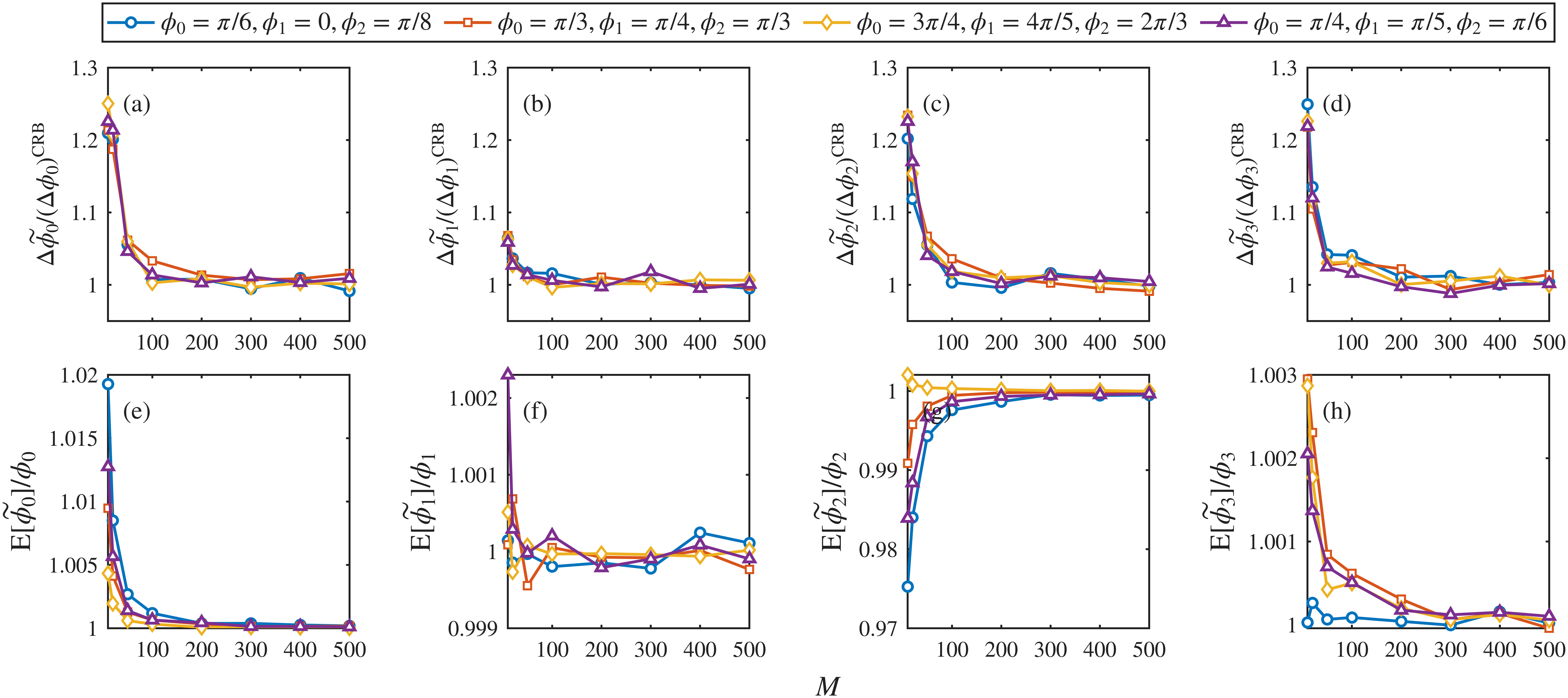}
    \caption{Asymptotic performance of the MLEs for a total mean photon number $N=10$ with equal resource split, $N_c = N_s = 5$. The different parameter values ($\phi_0, \phi_1, \phi_2$) in the figure assess the robustness of the estimation to changes in the true parameter values. Top row: the normalized CRB $\Delta\widetilde{\phi}_i/(\Delta\phi_i)^{\mathrm{CRB}}$ versus the number of experimental iterations $M$ for $i=0,1,2,3$. The convergence of all curves to unity shows that the MLE saturates the CRB in the large-$M$ limit. Bottom row: the unbiasedness ratio $\mathrm{E}[\widetilde{\phi}_i]/\phi_i$ as a function of $M$ for the same settings, which converges towards 1 as the estimator bias becomes negligible, typically for sample sizes of order $M\approx100$. The beam-splitter angle is set to the balanced working point $k_3=0$ ($\phi_3=\pi/4$), and the LO constants are fixed to $k_1=k_2=0.5$.}
    \label{fig:CRB_Bias_M_combined}
\end{figure}
The likelihood function for \(\bm{\phi}\) is given by
\begin{equation}
\mathcal{L}(\bm{\phi}\,|\,\bm{x}_1,\ldots,\bm{x}_M)
=\prod_{m=1}^{M} p(\bm{x}_m\,|\,\bm{\phi}),
\label{eq:App_Likelihood}
\end{equation}
and the MLE \(\widetilde{\bm{\phi}}_{\mathrm{MLE}}\) maximizes \(\mathcal{L}\), equivalently the log-likelihood \(\ell(\bm{\phi})=\log\mathcal{L}=\sum_{m=1}^{M}\log p(\bm{x}_m\,|\,\bm{\phi})\).
The estimator is obtained from the condition
\begin{equation}
\nabla_{\bm{\phi}}\ell(\bm{\phi})\Big|_{\bm{\phi}=\widetilde{\bm{\phi}}_{\mathrm{MLE}}}
=\bm{0}.
\label{eq:App_score_condition}
\end{equation}
Using \(\nabla_{\bm{\phi}}\log(\det\Sigma_{\bm{\phi}})=\mathrm{Tr}\!\left(\Sigma_{\bm{\phi}}^{-1}\nabla_{\bm{\phi}}\Sigma_{\bm{\phi}}\right)\) and rewriting the quadratic form in trace notation,
\[
(\bm{x}-\bm{\mu}_{\bm{\phi}})^{\mathsf T}\Sigma_{\bm{\phi}}^{-1}(\bm{x}-\bm{\mu}_{\bm{\phi}})
=
\mathrm{Tr}\!\left[\Sigma_{\bm{\phi}}^{-1}(\bm{x}-\bm{\mu}_{\bm{\phi}})(\bm{x}-\bm{\mu}_{\bm{\phi}})^{\mathsf T}\right],
\]
one arrives at the coupled MLE equations in the compact form
\begin{align}
\bm{0}
&=
\big(\nabla_{\bm{\phi}}\bm{\mu}_{\bm{\phi}}\big)^{\mathsf T}\,
\Sigma_{\bm{\phi}}^{-1}
\left(\bm{\mu}_{\bm{\phi}}-\frac{1}{M}\sum_{m=1}^{M}\bm{x}_m
\right)
+\frac{1}{2}\,
\mathrm{Tr}\!\left[
\big(\nabla_{\bm{\phi}}\Sigma_{\bm{\phi}}^{-1}\big)
\left(\Sigma_{\bm{\phi}}
-\frac{1}{M}\sum_{m=1}^{M}
(\bm{x}_m-\bm{\mu}_{\bm{\phi}})
(\bm{x}_m-\bm{\mu}_{\bm{\phi}})^{\mathsf T}
\right)
\right]
\;\;\Bigg|_{\bm{\phi}=\widetilde{\bm{\phi}}_{\mathrm{MLE}}}.
\label{eq:App_score_final}
\end{align}
Equation~\eqref{eq:App_score_final} is the standard score condition for a multivariate Gaussian model. In general, it does not admit a closed-form solution. In the main text, we solve it numerically and use the resulting MLE to verify the convergence of CRBs (Figs.~\ref{fig:CRB_vs_M} and~\ref{fig:CRB_vs_N} in the main text).

A useful simplification follows from the structure of the homodyne output statistics: in our scheme $\Sigma_{\bm{\phi}}$ is independent of the phase $\phi_1$. Hence $\partial_{\phi_1}\Sigma_{\bm{\phi}}=\bm{0}$, so the covariance-dependent trace term in Eq.~\eqref{eq:App_score_final} vanishes identically in the $\phi_1$ component. The MLE for $\phi_1$ therefore reduces to,
\begin{equation}
\left[(\partial_{\phi_1}\bm{\mu}_{\bm{\phi}})^{\mathsf T}\Sigma_{\bm{\phi}}^{-1}
\left(\bm{\mu}_{\bm{\phi}}-\frac{1}{M}\sum_{m=1}^{M}\bm{x}_m
\right)\right]_{\phi_1=\widetilde{\phi}_{1\mathrm{MLE}}}
=0,
\label{eq:App_phi_ML}
\end{equation}
where \(\widetilde{\bm{\mu}}_{\bm{\phi}}=\frac{1}{M}\sum_{m=1}^{M}\bm{x}_m\) is the estimator of  the sample mean \(\bm{\mu}_{\bm{\phi}}\). This makes explicit that the estimator \(\widetilde{\phi}_1\) is determined solely by the mean homodyne response. By contrast, the remaining parameters \(\phi_0\), \(\phi_2\), and \(\phi_3\) enter both \(\bm{\mu}_{\bm{\phi}}\) and \(\Sigma_{\bm{\phi}}\), and their MLEs follow from the equation \eqref{eq:App_score_final}. Figure~\ref{fig:CRB_Bias_M_combined} illustrates the asymptotic performance of the MLEs for different choices of the true parameter values. The results show that the estimators converge to the CRB and become effectively unbiased as the number of experimental iterations $M$ increases, indicating that the estimation scheme remains robust under variations of the underlying parameters.

\section{Effect of homodyne inefficiency}\label{Appendix D}
In this section, we quantify how finite homodyne inefficiency affects the FIM and corresponding CRBs. Our scheme saturates the CRBs simultaneously already for modest photon numbers, $N=5$--$10$ (see Fig.~\ref{fig:CRB_vs_N} in the main text). We identify the finite-photon-number regime in which the Heisenberg-scaling behavior remains visible for finite homodyne efficiencies.

\begin{figure}
    \centering
    \includegraphics[width=0.6\linewidth]{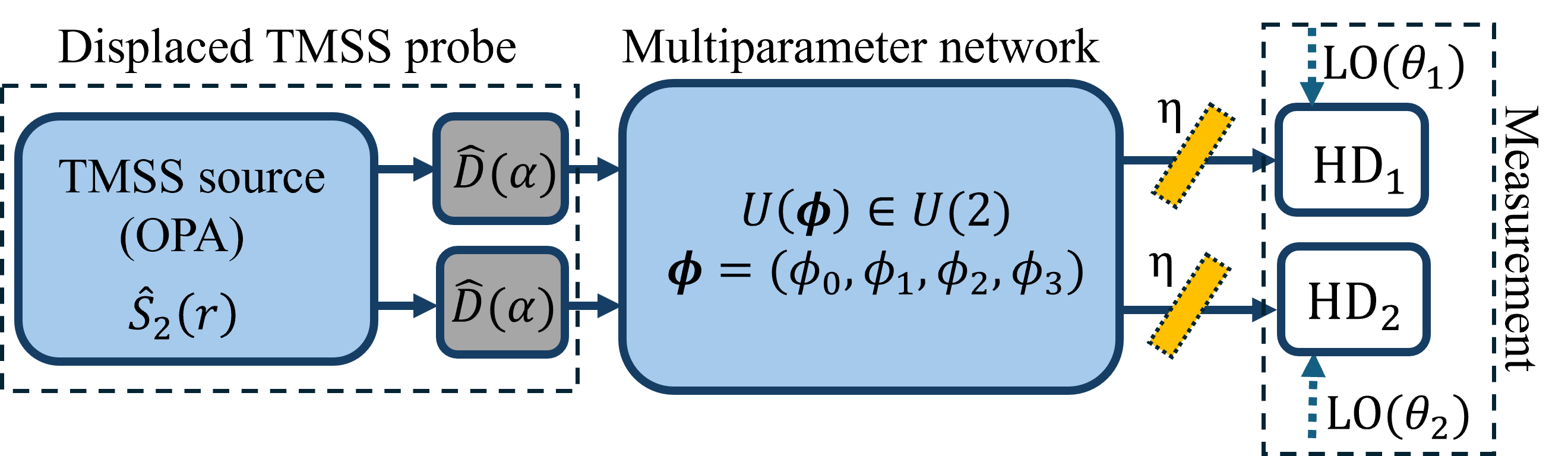}
    \caption{Detector inefficiency at each output port is modeled by a fictitious beam splitter of transmittance \(\eta\in(0,1]\), placed before the ideal balanced homodyne detectors at both output ports.}
    \label{fig:lossy_homodyne_model}
\end{figure}

In Fig.~\ref{fig:lossy_homodyne_model}, we model imperfect homodyne detection by a beam splitter of transmittance $\eta\in(0,1]$, preceding each ideal detector, which mixes the output modes with vacuum before detection~\cite{PhysRevA.96.062304}. For the two-port homodyne with equal efficiency $\eta$, the measured quadrature operator becomes $\hat{\bm{x}}_{\bm{\theta},\eta}=\sqrt{\eta}\,\hat{\bm{x}}_{\bm \theta} +\sqrt{1-\eta}\,\hat{\bm x}_{\rm vac}$, where $\hat{\bm x}_{\rm vac}$ has covariance $\mathbb I_2/2$, where $\mathbb{I}_2$ is $2\times2$ identity matrix. The joint two-port homodyne statistics remain Gaussian with~\cite{PhysRevA.96.062304, PRXQuantum.3.010202}
\begin{equation}
\bm{\mu}_{\bm{\phi}}^{\eta}
=
\sqrt{\eta}\,
\bm{\mu}_{\bm{\phi}},
\qquad
\Sigma_{\bm{\phi}}^{\eta}
=
\eta\,\Sigma_{\boldsymbol{\phi}}
+\frac{1-\eta}{2}\,\mathbb{I}_2.
\label{eq:lossy_moments}
\end{equation}
Since $\partial_i\boldsymbol{\mu}_{\boldsymbol{\phi}}^{\eta}=\sqrt{\eta}\,\partial_i\boldsymbol{\mu}_{\boldsymbol{\phi}}$ and $ \partial_i\Sigma_{\boldsymbol{\phi}}^{\eta}=\eta\,\partial_i\Sigma_{\boldsymbol{\phi}}$, from Eq.~\eqref{eq:FIMGaussian_main} in the main text, the inefficient-homodyne FIM can be written as
\begin{equation}
F_{ij}^{\eta}
=
\frac{\eta^2}{2}
\mathrm{Tr}
\left[
\left(\Sigma_{\boldsymbol{\phi}}^{\eta}\right)^{-1}
\left(\partial_i\Sigma_{\boldsymbol{\phi}}\right)
\left(\Sigma_{\boldsymbol{\phi}}^{\eta}\right)^{-1}
\left(\partial_j\Sigma_{\boldsymbol{\phi}}\right)
\right]
+
\eta\,
\left(\partial_i\boldsymbol{\mu}_{\boldsymbol{\phi}}\right)^T
\left(\Sigma_{\boldsymbol{\phi}}^{\eta}\right)^{-1}
\left(\partial_j\boldsymbol{\mu}_{\boldsymbol{\phi}}\right),
\label{eq:lossy_fim_compact}
\end{equation}
with $F^{\eta}=F^{\Sigma,\eta}+F^{\bm{\mu},\eta}$ and $(\Sigma_{\bm \phi}^{\eta})^{-1}$ is the inverse matrix of $\Sigma^\eta_{\bm \phi}$.

Under the LO-tuning condition in Eq.~\eqref{Eq:LOtuning} of the main text and at the balanced working point $\phi_3=\pi/4+k_3/N_s$, the ideal covariance matrix takes the asymptotic form given in Eq.~\eqref{App:AsymSigma}
\begin{equation}
    \Sigma_{\boldsymbol{\phi}}
=\frac{1}{N_s}\begin{pmatrix}
\frac{1+4k_1^2+4k_3^2}{4}& -(k_1+k_2)k_3\\
-(k_1+k_2)k_3 & \frac{1+4k_2^2+4k_3^2}{4}
\end{pmatrix}+O(1/N_s^{2}).
\end{equation}
The following analysis can be performed for arbitrary \(k_1,k_2,k_3\) except for the values in~\eqref{eq:Nonsingular_k}. For simplicity, we now take $k_1=k_2=1/2, k_3=0$, and the equal resource split $\beta=1/2$ used in the main text. The simplified covariance matrix $\Sigma_{\bm \phi}=\frac{1}{N}\mathbb{I}_2+O(1/N^2)$. The lossy covariance matrix~\eqref{eq:lossy_moments} then becomes
\begin{equation}
    \Sigma_{\bm{\phi}}^{\eta}=
\frac{\eta}{N}\mathbb{I}_2+\frac{1-\eta}{2}\mathbb{I}_2+O(1/N^{2})=\frac{L_N}{2}\,\mathbb{I}_2+O(1/N^2),
\label{eq:lossy_covariance_C}
\end{equation}
where $L_N=1-\eta+2\eta/N$. The inverse of the lossy covariance can be written as $(\Sigma^{\eta}_{\bm \phi})^{-1}=\frac{2}{L_N}\mathbb{I}_2+O(1)$. 
The first term is the covariance matrix in Eq.~\eqref{eq:lossy_covariance_C} is responsible for the \(N^2\) scaling of the FIM, while the second term is the vacuum noise introduced by detector inefficiency.
Substituting $(\Sigma_{\bm \phi}^{\eta})^{-1}$ and $\Sigma_{\bm \phi}^{\eta}$ into Eq.~\eqref{eq:lossy_fim_compact} then gives the leading inefficient-homodyne FIM,
\begin{equation}
F^{\eta}
\simeq
\mathrm{diag}
\left(
\frac{4\eta^2}{L_N^2},
\frac{\eta N}{2L_N},
\frac{\eta^2}{L_N^2},
\frac{4\eta^2}{L_N^2}
\right),
\qquad
L_N=
1-\eta+\frac{2\eta}{N}.
\label{eq:lossy_fim_representative}
\end{equation}
The corresponding CRBs are
\begin{equation}
\Delta^2\widetilde{\phi}_0^\eta
\geq
\frac{L_N^2}{4M\eta^2},
\qquad
\Delta^2\widetilde{\phi}_1^\eta
\geq
\frac{2L_N}{M\eta N},\qquad \Delta^2\widetilde{\phi}_2^\eta
\geq
\frac{L_N^2}{M\eta^2},
\qquad
\Delta^2\widetilde{\phi}_3^\eta
\geq
\frac{L_N^2}{4M\eta^2}
\label{eq:lossy_CRB}
\end{equation}
For the lossless detectors, $\eta=1$, Eq.~\eqref{eq:lossy_CRB} reduces to $\Delta^2\widetilde{\phi}_0^{\eta=1}
\geq
\frac{1}{M\,N^2},\;
\Delta^2\widetilde{\phi}_1^{\eta=1}
\geq
\frac{4}{M\,N^2},\; \Delta^2\widetilde{\phi}_2^{\eta=1}
\geq
\frac{4}{M\,N^2},
\;$ and $
\Delta^2\widetilde{\phi}_3^{\eta=1}
\geq
\frac{1}{M\,N^2}$ as evaluated in the main text. 

\begin{figure}
    \centering
    \includegraphics[width=0.6\linewidth]{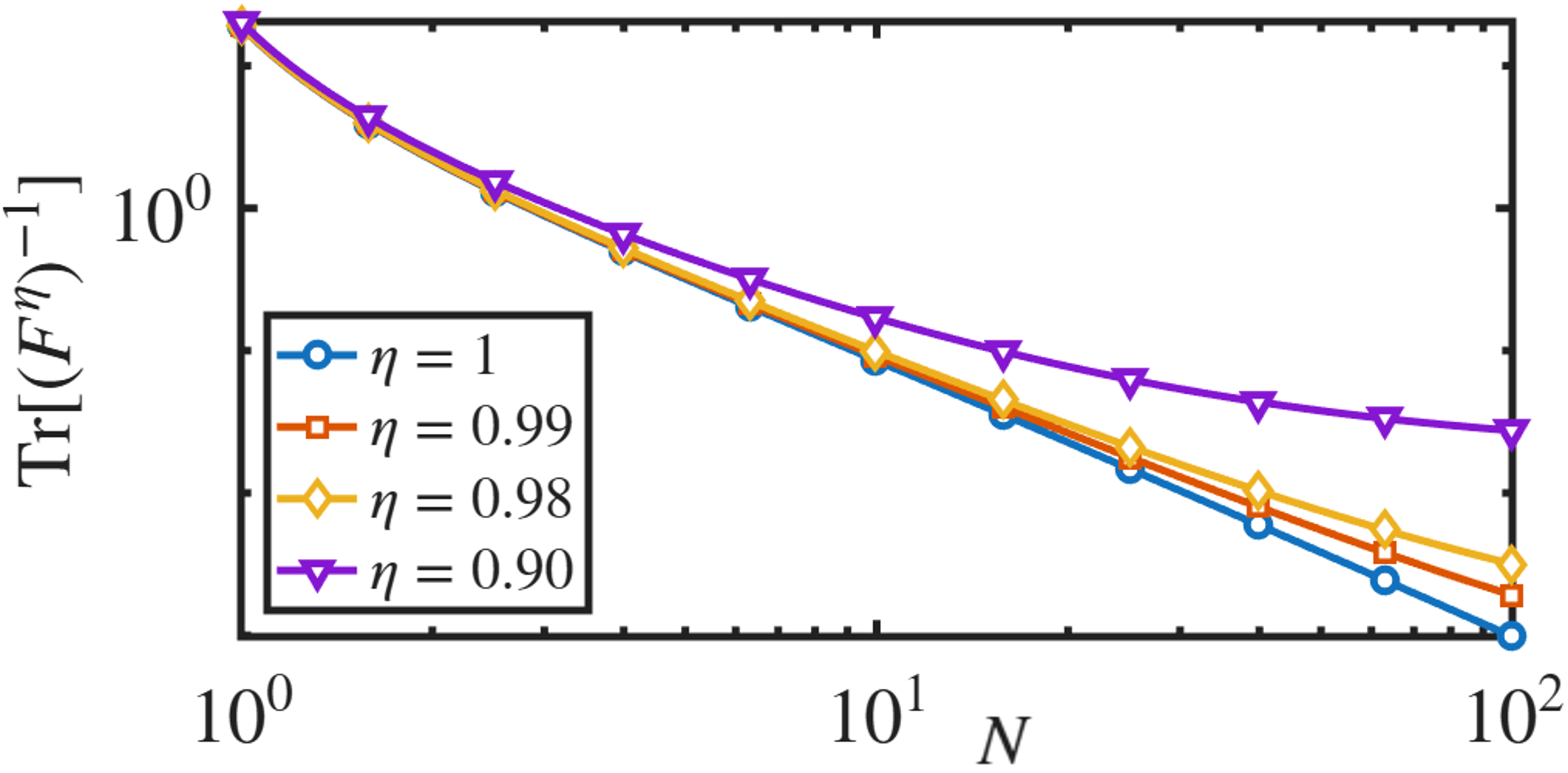}
  \caption{
 The log-log plot shows the scalar bound \(\mathrm{Tr}[(F^\eta)^{-1}]\), evaluated from the exact FIM in Eq.~\eqref{eq:lossy_fim_compact} under finite homodyne efficiency. The scalar bound remains close to the ideal case ($\eta=1$) for small detector inefficiencies in the low-photon-number regime. Here, $k_1=k_2=0.5$ and $k_3=0$.
}
    \label{fig:lossy_Scalar_bound}
\end{figure}

It is useful to define the finite-efficiency scale $N_\eta=2\eta/(1-\eta)$. We then analyze the regime $N\ll N_\eta$ for which the Heisenberg scaling is preserved. Defining
\begin{equation}
x=
\frac{N}{N_\eta}
=
\frac{N(1-\eta)}{2\eta},
\qquad
L_N=
\frac{2\eta}{N}(1+x),
\label{eq:r_loss_representative}
\end{equation}
The CRBs in Eq.~\eqref{eq:lossy_CRB} become
\begin{equation}
\Delta^2\widetilde{\phi}_0^\eta
\geq
\frac{(1+x)^2}{M\,N^2},
\qquad
\Delta^2\widetilde{\phi}_1^\eta
\geq
\frac{4 (1+x)}{M\, N^2},\qquad \Delta^2\widetilde{\phi}_2^\eta
\geq
\frac{4(1+x)^2}{M\,N^2},
\qquad
\Delta^2\widetilde{\phi}_3^\eta
\geq
\frac{(1+x)^2}{M\,N^2}.
\label{eq:lossy_HS_CRB_23}
\end{equation}
The condition $N\ll N_\eta$, equivalently $x\ll 1$, defines the regime in which all four CRBs retain the Heisenberg scaling with only small loss-dependent prefactor corrections.
The CRBs for parameters \(\phi_0,\phi_2,\phi_3\) acquire the factor $(1+x)^2$, whereas CRB of parameter $\phi_1$ acquires only the factor $(1+x)$. 
Equivalently, the degradation of Heisenberg-scaling precision is evaluated as the CRB uncertainty relative to the ideal detectors case
\begin{equation}
\frac{(\Delta\phi_0^{\eta})^{\rm CRB}}{(\Delta\phi_0^{\rm ideal})^{\rm CRB}}
=
\frac{(\Delta\phi_2^{\eta})^{\rm CRB}}{(\Delta\phi_2^{\rm ideal})^{\rm CRB}}
=
\frac{(\Delta\phi_3^{\eta})^{\rm CRB}}{(\Delta\phi_3^{\rm ideal})^{\rm CRB}}
=
(1+x),
\label{eq:loss_degradation_cov}
\end{equation}
whereas
\begin{equation}
\frac{(\Delta\phi_1^{\eta})^{\rm CRB}}{(\Delta\phi_1^{\rm ideal})^{\rm CRB}}
=
\sqrt{1+x}.
\label{eq:loss_degradation_mean}
\end{equation}
Thus, detector inefficiency affects the fluctuation-induced parameters more strongly than the parameter determined by the mean homodyne response. For \(1\%\) homodyne detection inefficiency and \(N=5\), the CRB uncertainties increases by only about \(2.5\%\) for \(\phi_0,\phi_2,\phi_3\), and by about \(1.3\%\) for \(\phi_1\). For $N=10$ and $1\%$ inefficiency the uncertainty is increased by about $5\%$ for \(\phi_0,\phi_2,\phi_3\) and by about $2.5\%$ for $\phi_1$. For \(2\%\) detection inefficiency, the increase remains moderate in the same photon-number range. 
Furthermore, Fig.~\ref{fig:lossy_Scalar_bound} shows the scalar CRB \(\mathrm{Tr}[(F^\eta)^{-1}]\) evaluated from the exact FIM in Eq.~\eqref{eq:lossy_fim_compact}. For high homodyne efficiencies, \(\eta=0.99\) and \(0.98\), the exact scalar bound remains close to the ideal result of Heisenberg-scaling behavior, only with a small correction in the prefactor, already for a few mean photons \(N\sim10\) used in the MLE analysis.
\end{document}